\documentclass[twocolumn,twoside,preprintnumbers,amsmath,amssymb,showkeys]{revtex4}

\usepackage{epsfig}

\usepackage{graphicx}

\usepackage{fancyhdr}
\usepackage{pslatex}

\begin{document}
\title {Non-identical particle femtoscopy in models with single freeze-out}

\author{Adam Kisiel$^1$ }

\affiliation{$^1$ Faculty of Physics,
		Warsaw University of Technology,
	ul. Koszykowa 75,
	00-662 Warsaw, Poland
}

\begin{abstract}

We present femtoscopic results from
hydrodynamics-inspired thermal models with single freeze-out. Non-identical
particle femtoscopy is studied and compared to results of identical
particle correlations. Special emphasis is put on shifts  
between average space-time emission points of non-identical particles
of different masses. They are found to 
be sensitive to both the spatial shift coming from radial flow, as
well as average emission time difference coming from the resonance
decays. The Therminator Monte-Carlo program was chosen for this study
because it realistically 
models both of these effects. In order to analyze the results we
present and test the methodology of 
non-identical particle correlations.

\end{abstract}

\vskip -1.35cm

\maketitle

\thispagestyle{fancy}

\setcounter{page}{1}

\bigskip

\section{Introduction}

The single freeze-out
approach~\cite{Florkowski:2001fp,Broniowski:2001we,Broniowski:2002wp}
originates from thermal models of the
heavy-ion collisions. It is based on thermal fits to particle yields
and yield ratios. Since these fits are known to
work well for RHIC collisions, the single freeze-out model is also reproducing them
correctly. These ratios are not sensitive to the underlying geometry
of the collision, which is what is measured by femtoscopy. The
form of the freeze-out geometry must be postulated, and should give the
overall volume of the system, which is reflected in absolute yields of
particles, as well as detailed shape of the emission region which can
be probed by two-particle correlations. We have postulated such a form of
the freeze-out hypersurface which is motivated by hydrodynamics. It
has been used to calculate femtoscopic observables, both for identical~\cite{Kisiel:2006is}
and non-identical particles. The latter are especially
interesting and are the focus of this work. They have been recently measured at
RHIC~\cite{Adams:2003qa}. There have been very few theoretical 
predictions for these observables~\cite{LisaRetiere}. They contain a crucial and unique
piece of information - the difference between the average emission
points of two particle
types~\cite{Lednicky:1982,Lednicky:1995vk,Voloshin:1997jh,Lednicky:2005tb,Kisiel:2006si}. If the
particles have different masses, 
we expect a spatial shift coming from collective flow of matter. If
the particles are of different type, we expect very different pattern
of emission times, since many particles come from strong decays of
resonances. Both of these shifts are interconnected in the measured
average emission point difference. Disentangling them is not a trivial
task. Single freeze-out models with resonances are perfectly suited
for it, as they include realistic modelling of both effects.

\section{Femtoscopy definitions}

In this work we will analyze correlation functions between
non-identical particles.
Femtoscopic correlation function is usually defined as:
\begin{equation}
\label{cfgen}
C(\vec q, \vec K) = \frac{P_{2}^{C}(\vec q, \vec K)}{P_{2}^{0}(\vec q, \vec K)}
\end{equation}
where $P_2^{C}$ is the probability to observe two femtoscopically
correlated particles at relative momentum $\vec q$. $P_2^{0}$ is such
probability where the correlation between particles does not have the
femtoscopic component. $\vec K$ is the average momentum of the
pair. In heavy-ion experiments the $P_2^{C}$ is usually constructed
from pairs coming from the same event; $P_2^{0}$ from pairs where each
particle comes from a different event, but the events are as close to
each other in global characteristics as possible.

In theoretical models one should, in principle, generate particles
in such a way that they are already correlated due to their mutual and
many-particle interactions. That is however usually computationally
not possible. One then makes an assumption that the interaction
between particles can be separated from the generation process and we write
the most general form of the correlation function that can be used by
models:
\begin{equation}
\label{cftheory}
C(\vec q, \vec K) = \frac{\int S_{1,2}(\bold r^{*}, \vec q, \vec K) \left|
\Psi(\vec q, \bold r^{*}) \right|^2 d^{4} \bold r^{*}}{\int S_{1,2}(\bold r^{*}, \vec q, \vec K) d^{4} \bold r^{*}}
\end{equation}
where $\bold r^{*}$ is the pair separation in the pair rest frame
(PRF) and $S_{1,2}(\bold r^{*}, \vec q, \vec K)$ is the pair separation
distribution defined as:
\begin{equation}
\label{Sofr}
S_{1,2}(\bold r^{*}, \vec q, \vec K) = \int S_{1}(\bold x_{1}, \vec p_{1}) S_{2}(\bold
r^{*} - \bold x_{2}, \vec p_2) \delta(\bold r^{*} - \bold x_{1} +
\bold x_{2}) d^4 \bold x_{1} d^4 \bold x_2 
\end{equation}
and $S(\bold x, \vec p)$ is the single-particle emission function
provided by the model. For identical particles $S_1 \equiv S_2$ and
$S_{1,2}(\bold r^{*})$ is symmetric by definition, for non-identical
particles it is not so and $S_{1,2}(\bold r^{*})$ is usually asymmetric. It
is an important point which will be discussed later. We also note that
the ordering of particles in the pair is important: $S_{1,2}(\bold r) \equiv 
S_{2,1}(-\bold r)$.

\subsection{Wave-function of the pair}

In this work we consider pairs of charged non-identical mesons and baryons. In
such case the origin of femtoscopic correlations are Coulomb and
strong interactions. However for pion-kaon and pion-proton systems, as
well as for same-charge kaon-proton system, the strong
interaction is much smaller than Coulomb. For opposite charge
kaon-proton system, there is a significant strong potential, which is
interesting in it's own right, its detailed study is however beyond
the scope of this paper. The strong interaction is not
essential for our study, so we restrict our study to Coulomb
interaction in same-charged pion-kaon, pion-proton and
kaon-proton systems. The pair wave-function squared is then:
\begin{eqnarray}
\label{psiqc}
\left| \Psi(\vec q, \bold r^{*})^{QC} \right|^2 &=& A_C |
\exp(-i \vec k^{*} \vec
r^{*})F(-i \eta, 1, i \xi) 
|^2
\end{eqnarray}
where $k^{*}$ is half of pair relative momentum in PRF, $A_{C}$ is the
Gamow factor, $F$ is the confluent hypergeometric function, $\eta =
1/k^{*} a_{c}$, $a_c$ is the pair Bohr radius and $\xi = k^{*}
r^{*} + \vec k^{*} \vec r^{*}$. Please note that the wave-function is
calculated in the PRF. $a_c$ is $248.5 fm$, $222.6 fm$ and $83.6 fm$
for pion-kaon, pion-proton and kaon-proton pair respectively.

\section{Non-identical particle correlations}
\label{nonidcorr}

The correlation between a pair of non-identical particles arises from
Coulomb and/or strong interaction. We will
concentrate on the Coulomb interaction, but our conclusions hold for
strong interaction as well. We will discuss the specifics of the
correlations for pairs of unlike particles, emphasizing the
differences and similarities to traditional identical particle
femtoscopy.  

The wave-function (\ref{psiqc}) is calculated in the pair rest frame
and depends on relative momentum $k^{*}$, relative position $r^{*}$
and the angle $\theta^{*}$ between the two. It needs to be emphasized
that low relative momentum in pair rest frame, corresponds to close
{\it velocities}, but not momenta, in the laboratory frame. This is in
contrast to the identical particle interferometry. This also means
that particles from very different momentum ranges are correlated,
e.g. pion with velocity $0.7$ has a momentum of $0.137 GeV$, a close-velocity kaon
has a momentum of $0.484 GeV$ and a proton: $0.919 GeV$. In experiment
this often poses a problem, as one needs to have a large momentum
acceptance to measure close-velocity pairs. 

Looking in detail at the hypergeometric function $F$ from
(\ref{psiqc}):
\begin{equation}
F = 1 + r^{*}(1+\cos \theta^{*})/a + ...
\label{fdecomp}
\end{equation}
one notices an important feature of the wave-function, namely that it
is not symmetric with respect to the sign of $\cos \theta^{*}$. For
same sign particles $A_c$ is less than 1.0 and $F$ is above 1.0, but
since the correlation effect must be negative, $1-A_c > F-1$. For a
given $k^{*}$ and $r^{*}$ one can have two cases: in one $\cos
\theta^{*}<0$, for the other $\cos \theta^{*}>0$. The former will
have a {\it larger} correlation effect (since $F$ is smaller and
cannot overcome $A_c$) and the latter will have a {\it smaller}
correlation effect. This asymmetry in the correlation effect can be
understood with a help of a simple picture: negative $\cos \theta^{*}$
means that $k^{*}$ and $r^{*}$ are anti-aligned, which means that at
first the particles will fly towards each other before they fly away, spending
more time close together and thus developing a larger correlation. A
positive $\cos \theta^{*}$ means they will fly away immediately,
having no time to interact. This asymmetry is an intrinsic property of
the Coulomb interaction and is also present for identical
particles. However in that case the wave-function symmetrization
requires one to add a second term to (\ref{psiqc}) which has the same
asymmetry with opposite sign, so that the overall asymmetry is zero,
as it must be.

If we were somehow able to divide pairs in
two groups - one in which $\left<\cos \theta^{*}\right> >0$ and the other in
which $\left<\cos \theta^{*}\right> <0$ one would obtain two different
correlation functions, out of which the latter would show a stronger
correlation effect. Obviously we cannot select pairs based on the
$\theta^{*}$ angle, as it is not measured. However we do have one
angle on which we can select - the angle $\Psi$ between pair velocity
$v$ and pair relative momentum $k^{*}$. In the transverse plane $\cos
\Psi>0$ means that $k^{*}_{out} >0$. We also notice that in the
transverse plane
\begin{equation}
\Psi = \theta^{*} + \phi
\label{anglesum}
\end{equation}
where $\phi$ is the angle between pair velocity $v$ and relative
position $r^{*}$. This angle is not measured as well. One can also show that
when one averages over all possible positions of $r^{*}$ one can
write~\cite{Kisiel:2004phd}
\begin{equation}
sign\left<\cos \Psi\right> = sign\left<\cos \theta^{*}\right> sign\left<\cos \phi\right>.
\label{angleav}
\end{equation}
One can then propose a measurement: to divide pairs into two groups, one
with $k^{*}_{out} > 0 \equiv \cos \Psi > 0$ and the other with
$k^{*}_{out}<0 \equiv \cos \Psi<0$. Then one constructs two
correlation functions $C_{+}$ and $C_{-}$ from the two groups. If one
observes that $|C_{+}-1|/|C_{-}-1.0|>1.0$ it can only happen if $\left<\cos
\phi\right> < 0$ as can be seen from Eq. (\ref{angleav}). In other words
this can happen only if the average emission points of two particle
species are, on the average, separated in the direction of pair
velocity, and this separation is anti-aligned with the velocity. By
the same reasoning, if $|C_{+}-1|/|C_{-}-1.0|<1.0$ this separation is
aligned with velocity. Let us restate the conclusion of this
paragraph: using the fact that the correlation effect is asymmetric
with respect to the sign of $cos(\theta^{*})$ and the measured angle
$\Psi$ we can tell whether an average emission position of two different
particle species is the same or not, and if it is not, is the
difference in or opposite to the direction of pair velocity $v$. More
quantitative analysis shows, that the double ratio $C_{+}/C_{-}$ is
also monotonously dependent on the value of this shift between
particles, so the magnitude of the shift can be inferred from it. This
a unique feature of non-identical particle 
correlations, such information cannot be obtained from any other
measurement. 

The above consideration has been done for the pair rest
frame. Experimentally however, one would like to learn something about
the source itself, which requires the knowledge of the source in it's
rest frame. One can write a simple formula for the relative separation
in pair rest frame $r^{*}$ as a function of pair separation in source
frame $r$:
\begin{eqnarray}
r^{*}_{out} &=& \gamma_t (r_{out} + \beta_t \Delta t_L) \nonumber \\
r^{*}_{side} &=& r_{side} \nonumber \\
r^{*}_{long} &=& \gamma_l (r_{long} + \beta_l \Delta t) \nonumber \\
\Delta t_L &=& \gamma_l (\Delta t + \beta_l r_{long})
\label{rstarofr}
\end{eqnarray}
where the $long$ direction is defined as the one parallel to the
velocity of the colliding nuclei,
$out$ as parallel to the pair velocity in the transverse direction,
and $side$ as perpendicular to the other two. Pair velocities $\beta_l = p_l/E$,
$\gamma_l = 1/\sqrt{1-\beta_l^2}$, $\beta_t = p_{\perp}/m_{\perp}$,
$\gamma_t = 1/\sqrt{1-\beta_t^2}$.
One can see that the
average shift in $r^{*}_{out}$ may mean non-zero average spatial shift
$r_{out}$, non-zero average emission time difference $\Delta t$ or the
combination of the two. 

\section{THERMINATOR}
\label{sec:Therm}

The Therminator program~\cite{Kisiel:2005hn} is the numerical
implementation of the single freeze-out
model~\cite{Florkowski:2001fp,Broniowski:2001we,Broniowski:2002wp}. It
includes all 
particles listed by the Particle Data Group~\cite{Eidelman:2004wy}. We use the
version of the model based on blast-wave type
parametrization, which is hydrodynamics-inspired~\cite{LisaRetiere}. Therefore our
model includes the effects of radial flow, which is apparent e.g. in
the $m_T$ dependence of the pion ``HBT
radii''~\cite{Kisiel:2006is}. This is an important feature, as the
space-momentum correlation coming from radial flow is one of the
origins of the emission asymmetries between various particle species.
The emission function is changed slightly from the version described
in~\cite{Kisiel:2005hn}. A quasi-linear velocity profile as a function
of $\rho$ is added. The freeze-out hypersurface is then defined as:
\begin{equation}
\tilde{\tau} = \tau = const, ~~~~~v_r = \tanh \alpha_{
\perp}(\zeta) = \frac{\rho/\rho_{max}}{v_{T}+\rho/\rho_{max}}
\label{velproffreeze}
\end{equation}
where $\tau$, $\rho_{max}$ and $v_T$ are parameters of the model. The quasi-linear velocity
profile has the desired features - it is zero for $\rho = 0$, is
almost linear for reasonable values of $v_{T}$ and cannot go higher
than 1.0. The emission function is then:
\begin{eqnarray}
& &\frac{dN}{dy d\varphi p_{\perp} dp_{\perp} d\alpha_{\parallel} d\phi
\rho d\rho} = \frac{\tau}{(2 \pi)^3} m_{\perp}
\cosh(\alpha_{\parallel}-y) \times  
\label{Srtherm}
 \\
& &\left\{ \exp \left[ \beta \frac{
m_{\perp}\cosh(\alpha_{\parallel}-y) - p_{\perp} v_r \cos(\varphi -
\phi)}{\sqrt{1-v_r^2}} - \beta \mu \right]
\pm 1\right\} \nonumber
\end{eqnarray}

\begin{figure}[!htb1]
\begin{center}
\includegraphics*[angle=0, width=7cm]{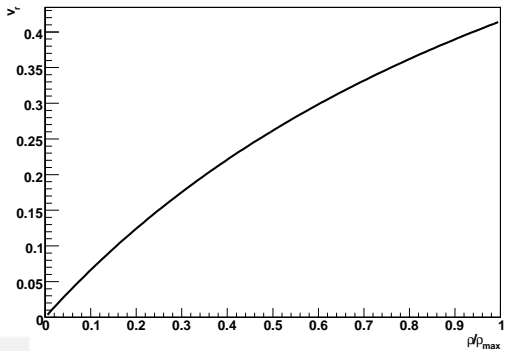}
\end{center}
\caption{\emph{ \small Radial velocity profile for the parameters used
in this study.}}
\label{fig:velprof}
\end{figure}

The fit to STAR Collaboration data~\cite{Adams:2003qm} has been performed
with this model and the following values of the parameters were
found to best reproduce the observed pion and kaon spectra in central AuAu
collisions: $\tau = 8.55 fm$, $\rho_{max} = 8.92 fm$, $a = -0.5$, $v_{T}
= 1.41$. The thermodynamic parameters were the same as
in~\cite{Kisiel:2006is}. The velocity profile for these parameters is
shown in Fig. \ref{fig:velprof}. The average velocity is $0.31$.

\begin{figure}[!htb1]
\begin{center}
\includegraphics*[angle=0, width=7cm]{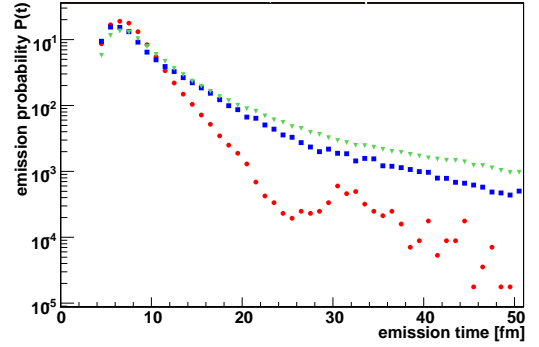}
\end{center}
\caption{\emph{ \small Probability to emit a pion (green triangles),
kaon (blue squares) and proton (red circles) as a function of time.}}
\label{fig:timedep}
\end{figure}

\begin{table}[htb]
\begin{center}
\caption{\bf Average emission times from Therminator}
\begin{tabular}{|l|l|}
\hline
Particle species & Average emission time \\
\hline\hline 
pion & 12.3 $fm/c$ \\
 \hline
kaon & 10.7 $fm/c$ \\
 \hline
proton & 7.9 $fm/c$ \\
 \hline
\end{tabular} \label{emtimes}
\end{center}
\end{table}

The simulation proceeds as follows. First, using a Monte-Carlo
integration procedure as a particle generator, all particles (stable
and unstable) are generated according to the emission function
(\ref{Srtherm}). Each particle is given an emission point on the
freeze-out hypersurface and a momentum. Then all unstable particles
decay after some random time  dependent on their width. They
propagate to the decay point, and this point is taken as the
space-time origin of the daughter particles. Two- and three-particle decays are
implemented. The process is repeated for cascade decays, until only
stable particles remain. While each particle has emission point on the
freeze-out hypersurface (we call such particles primordial) or at the
decay point of the heavier resonance, the full history of decays can
be reconstructed from the output files. This treatment of resonance
propagation and decay is of crucial importance for non-identical
particle correlations. It introduces delay in emission time, different
for different particle species. It depends on the number of resonances
that decay into the particle of interest, their widths and
velocities. An illustration can be seen on Fig.~\ref{fig:timedep}. The
Monte-Carlo procedure we used is the most efficient way 
of studying it, more precise results can only be obtained by using a
full-fledged hadronic rescattering model. In our work hadronic
rescattering is not taken into account, which is one of the
simplifications assumed in the single freeze-out model. 

\subsection{Calculating the correlation function}

It is possible to use Eq. \ref{cftheory} to get the model correlation
function. The integral is calculated numerically. One takes particles
generated by Therminator. Then one combines them 
into pairs and creates two histograms - in one of them one stores the
squared wave-function of the pair (\ref{psiqc}), in the other - unity for each
pair. The result of the division of the two histograms is the average
pair wave function in each bin, which is the correlation function
 per Eq. (\ref{cftheory}). This is the so-called
``two-particle weight'' method of calculating the correlation function
from models. It is the only one which enables to take into account the
two-particle Coulomb interaction exactly, a feature which is
necessary for our study. In the procedure each pair is treated
separately, for each of them one goes to the pair rest frame
(different for each pair) as the calculation of the pair wave function
is done most naturally in this system.

\begin{figure}[!htb1]
\begin{center}
\includegraphics*[angle=0, width=7cm]{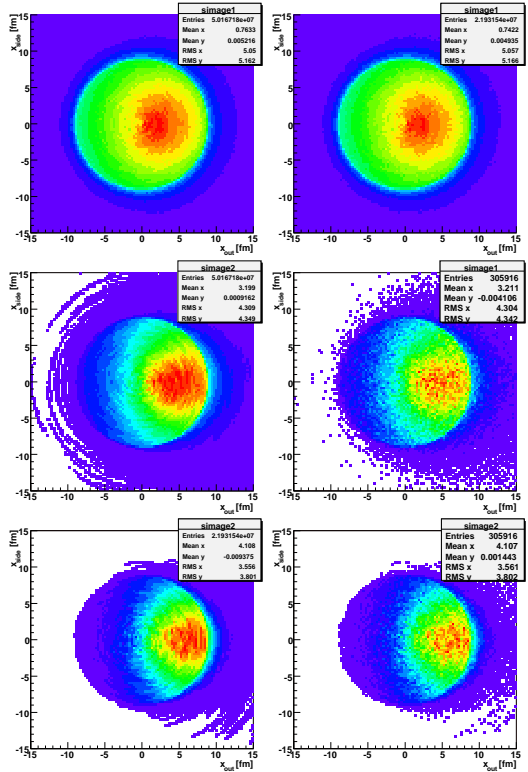}
\end{center}
\caption{\emph{ \small Distribution of emission points of pions (upper
plots), kaons (center plots) and protons (lower plots) versus
pair-wise out and side directions. Left and right side are emission
points of the same particle from different system, but the same pair
velocity range (for pions: left - $\pi K$, right - $\pi p$, for
kaons: left - $\pi K$, right - $K p$, for protons: left - $\pi p$,
right - $K p$}}
\label{fig:zrodelka}
\end{figure}

\section{Asymmetry analysis}

In section \ref{nonidcorr} it was shown that non-identical particle correlations
are sensitive to the shifts between average emission points of
different particle species. However we have not discussed if and how
such asymmetries could arise. In Fig.~\ref{fig:zrodelka} average emission
points of pions, kaons and protons from THERMINATOR where shown, for pairs which have
similar velocity, pointing horizontally to the right. One sees
right away that the average emission points of pions, kaons and
protons is {\it not} the same in the $out$ direction, while it is 0
for the side direction for all of them. One can also see that the
average size of the emission region is decreasing with particle mass,
a known effect of radial flow, usually referred to as the ``$m_T$
scaling'' of HBT radii. The shift that we observe is also a direct but
distinctly different consequence of radial flow present in our
simulation. One can understand it by the following argument. Close
velocity pions and protons have very different momenta. By inspecting
Eq. (\ref{Srtherm}) one observes that the correlation between
space-momentum emission point direction $\phi$ and momentum direction
$\varphi$ is controlled by the factor: $\exp(\beta p_{\perp}
\cos(\varphi - \phi))$. So the higher the momentum, the stronger the
correlation. Temperature $1/\beta$ is of course identical for both
particle species. Therefore, in a close-velocity pair, proton emission
direction will be much more correlated with its momentum
direction. And the momentum, due to radial flow, is always pointing
``outward'', so the emission points will tend to be concentrated near
the edge of the system, in the direction of emission. For pions on the
other hand, there is almost no correlation between the two directions,
so they are emitted from the whole source. This is clearly seen in the
figure, as a difference between mean emission points in the ``out'' direction:
\begin{equation}
r^{\pi p}_{out} = x^{\pi}_{out} - x^{p}_{out}
\label{rout}
\end{equation}
This is the spatial shift between particles  in the source
frame. Measuring it is the main goal of non-identical particle
femtoscopy. Observing such shift in the experiment would be a direct
evidence of the collective behaviour of matter, which is one of the
necessary conditions to claim the discovery of the quark-gluon
plasma. 

One must remember that the asymmetries measured in the correlation
function are averaged over the source. Therefore the connection
between shift in laboratory frame and pair rest frame is:
\begin{equation}
\left< r^{*}_{out} \right> = \left< \gamma \left(r_{out} - \beta
\Delta t \right) \right>
\label{avrstar}
\end{equation}

\begin{table}[htb]
\begin{center}
\caption{\bf Average space shifts from Therminator}
\begin{tabular}{|l|l|l|l|}
\hline
$\beta_t$ of the pair & $r_{\pi K}$ & $r_{\pi p}$ & $r_{K p}$  \\
\hline\hline 
0.35 - 0.5 &  -1.9 fm & -2.5 fm & -0.6 fm\\
 \hline
0.5 - 0.65 &  -2.4 fm & -3.4 fm & -0.9 fm\\
 \hline
0.65 - 0.8 &  -2.9 fm & -3.9 fm & -1.1 fm\\
 \hline
0.8 - 0.95 &  -3.0 fm & -4.3 fm & -1.2 fm\\
 \hline
\end{tabular} \label{spacediffs}
\end{center}
\end{table}

\begin{table}[htb]
\begin{center}
\caption{\bf Average time shifts from Therminator}
\begin{tabular}{|l|l|l|l|}
\hline
$\beta_t$ of the pair & $\Delta t_{\pi K}$ & $\Delta t_{\pi p}$ &
$\Delta t_{K p}$  \\
\hline\hline 
0.35 - 0.5 &  3.8 fm/c & 6.2 fm/c & 2.3 fm/c\\
 \hline
0.5 - 0.65 &  3.6 fm/c & 5.7 fm/c & 2.1 fm/c\\
 \hline
0.65 - 0.8 &  3.2 fm/c & 5.1 fm/c & 1.9 fm/c\\
 \hline
0.8 - 0.95 &  2.6 fm/c & 4.2 fm/c & 1.5 fm/c\\
 \hline
\end{tabular} \label{timediffs}
\end{center}
\end{table}

\begin{figure}[!htb1]
\begin{center}
\includegraphics*[angle=0, width=7cm]{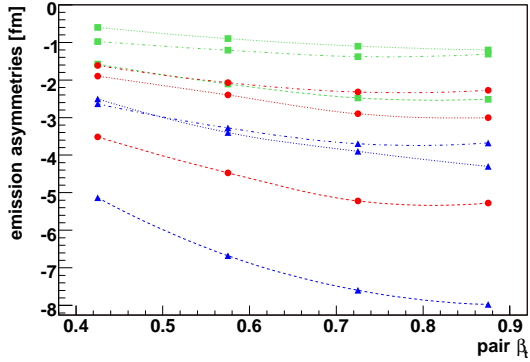}
\end{center}
\caption{\emph{ \small Components of the average shift between
particle species in the laboratory frame. Dotted line is space
component, dash-dotted line is time 
component, dashed line is full asymmetry. Red circles is $\pi K$
system, blue triangles is $\pi p$, green squares is $K p$. }}
\label{fig:avcomp}
\end{figure}

From Eq. (\ref{rstarofr}) one sees that the measurable shift
$r^{*}_{out}$ is a combination of spatial shift $r_{out}$, and
emission time shift $\Delta t$. The former effect is of special interest and has
been studied in detail in \cite{LisaRetiere}. However the effect of
time difference has not been adequately studied so far. The
Therminator model has been chosen to perform this task, because it
includes both effects in the same calculation in a self-consistent
way. Fig.~\ref{fig:timedep} shows the probability to emit a pion, a kaon
or a proton at a given time. One sees that average emission times of
this particle species, listed in Tab. \ref{emtimes}, are different in
the laboratory frame. This will affect the observed asymmetries. 

In Fig.~\ref{fig:avcomp} one can see the components of the emission
asymmetries in the laboratory frame, obtained directly from the
emission functions $S_{1,2}$. The time and space components for
all considered systems are compared. They are of the same order for
all systems. The overall asymmetry is the combination of the
two. 

\subsection{Obtaining femtoscopic information}

The correlation function for non-identical particles is given by
Eq. (\ref{cftheory}). In order to obtain femtoscopic information from
the experimental correlation function one needs to perform a fit
procedure. In traditional HBT measurement the integral analog to
Eq. (\ref{cftheory}) can be performed analytically to obtain a
simple fit function. It is not possible for the general case of
non-identical particles, where the following procedure must be applied. Usually
smoothness approximation is used, that is the emission function
factorizes into a space and momentum components. A form
of the spatial emission function is then assumed:
\begin{equation}
S(\vec r) \approx exp \left( -\frac{(r_{out} - \mu_{out})^2 +
r_{side}^2 + r_{long}^2}{2R^2} \right)
\label{sofx}
\end{equation}
defined in the laboratory frame. Then momenta of pairs of particles
are taken from the experiment. Their emission points are randomly
generated according to (\ref{sofx}), with some assumed values of
source parameters: the gaussian source radius $R$ and the shift in the
outwards direction $\mu_{out}$. For each pair the weight is then calculated
according to (\ref{psiqc}), and the average of the weights over all
pairs as a function of $k^{*}$ is constructed. This is a theoretical
correlation function according to (\ref{cftheory}). This function can
be compared via a $\chi^2$ test to the experimental correlation
function being fitted. By varying parameters $R$ and $\mu_{out}$ a
function can be found which best describes the input one. Parameters
of the source which produce this function are taken as the best-fit
values. In this way femtoscopic information is obtained from the
non-identical particle correlation function. In this work the
model correlation functions calculated according to Eq.~\ref{cftheory}
were treated as ``pseudo-experimental'' ones. A dedicated
program~\cite{Kisiel:2004phd,Kisiel:Nukl} has been used to perform a
fit on them in a manner as closely resembling the experimental
situation as possible. It was a direct test of the methodology of
non-identical particle correlations.

\subsection{Sum rule for shifts between different particle species}

If we have three different particle species, we have three average
emission point shifts that we can measure: $\mu^{\pi K}$, $\mu^{\pi
p}$ and $\mu^{K p}$. However, if we take the same group of e.g. pions
for $\pi-K$ correlations and $\pi-p$ correlations (and similarly the
same group of kaons and protons), we might expect that
a simple sum rule holds:
\begin{equation}
\mu^{\pi p} = \mu^{\pi K} + \mu^{K p}
\label{sumform}
\end{equation}
and only two of the shifts are independent. 

Non-identical particles are correlated if they have close
velocities. If we select pairs of particles with some pair velocity,
we expect that the correlated particles themselves also have
velocities in this range. Therefore if we compare pions that form the
$\pi-K$ correlation in the pair $\beta_t$ range between 0.5 and 0.65, and
pions from the $\pi-p$ correlation in the same pair $\beta_t$ range we may
assume that these are the same pions (providing that we
construct the correlation functions from the same events). That shows
that pair $\beta_t$ is the correct variable to select on, when studying
momentum dependence of the femtoscopic parameters from non-identical
particle correlations. It also shows that we should indeed expect the
sum rule (\ref{sumform}) to hold separately in each $\beta_t$ bin.  

\subsection{Two-particle versus single particle sizes}

The size of the two-particle source is a combination of the individual
single-particle source sizes:
\begin{eqnarray}
\sigma_{\pi K} &=& \sqrt{\sigma_{\pi}^2 + \sigma_{K}^{2}} \nonumber \\
\sigma_{\pi p} &=& \sqrt{\sigma_{\pi}^2 + \sigma_{p}^{2}} \nonumber \\
\sigma_{K p} &=& \sqrt{\sigma_{K}^2 + \sigma_{p}^{2}} 
\label{twofromone}
\end{eqnarray}
where $\sigma$ is a width of a gaussian fitted to the corresponding
emission function (either single- or two-particle). One can
immediately see that the sizes are not independent, similar to
shifts. One can also use the combination of the measured two-particle
source sizes to obtain the single particle sizes:
\begin{eqnarray}
\sigma_{\pi} &=& \sqrt{\frac{\sigma_{\pi K}^2 + \sigma_{\pi p}^{2} - \sigma_{K p}^{2}}{2}} \nonumber \\
\sigma_{K}   &=& \sqrt{\frac{\sigma_{\pi K}^2 - \sigma_{\pi p}^{2} + \sigma_{K p}^{2}}{2}} \nonumber \\
\sigma_{p}   &=& \sqrt{\frac{-\sigma_{\pi K}^2 + \sigma_{\pi p}^{2} + \sigma_{K p}^{2}}{2}} 
\label{onefromtwo}
\end{eqnarray}
They can then be compared to single-particle source sizes obtained
from regular identical particle interferometry to see whether the size
of the system is described self-consistently.

\section{Results and discussion}

\begin{figure}[!htb1]
\begin{center}
\includegraphics*[angle=0, width=7cm]{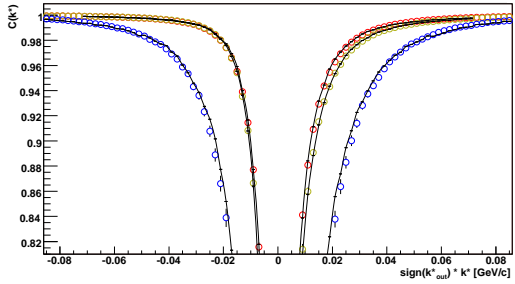}
\end{center}
\caption{\emph{ \small Correlation functions for pion-kaon (red),
pion-proton (yellow) and kaon-proton (blue) for pairs with velocity
between 0.5 and 0.65. The lines are fits to the correlation function.}}
\label{fig:pikpfun}
\end{figure}

The analysis of the correlation functions for pion-kaon, pion-proton
and kaon-proton systems has been performed based on the Therminator
model and the two particle weight method, described in the previous
paragraphs. Examples of the obtained correlation functions are shown
in Fig. \ref{fig:pikpfun}. As expected for like-sign pairs they go
below unity at low $k^{*}$. The correlation effect is the smallest for
pion-kaon and largest for kaon-proton due to the difference in the
Bohr radii of the pairs. It is a fortunate coincidence, as for
kaon-proton one expects to have the smallest statistics in the
experiment, but due to a large correlation effect the measurement should
be doable for a data sample similar to the one in which pion-kaon
measurement is possible.

\begin{figure}[!htb1]
\begin{center}
\includegraphics*[angle=0, width=7cm]{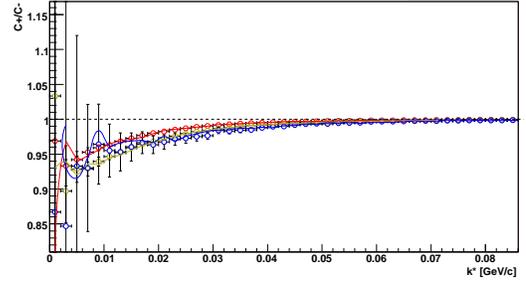}
\end{center}
\caption{\emph{ \small Double ratios for pion-kaon (red),
pion-proton (yellow) and kaon-proton (blue) for pairs with velocity
between 0.5 and 0.65. The lines are double ratios calculated from the
fits to the correlation function.}}
\label{fig:pikpdr}
\end{figure}

In Fig. \ref{fig:pikpdr} examples of the asymmetry measurement - the
double ratios $C_{+}/C_{-}$ for pion-kaon, pion-proton and kaon-proton
systems, are shown. One can see a significant signal, which indicates
that the correlation functions are indeed sensitive to the asymmetries
described in the previous paragraph. In our analysis we have adopted a
convention in which a lighter particle is always taken as first in the
pair.  All the double ratios go below unity, which means that, on
average, particles which are lighter are emitted closer to the center
of the system or later (or both). This is exactly consistent with the
qualitative picture of spatial shifts coming from radial flow shown in
Fig. \ref{fig:zrodelka} and with emission time differences coming
from resonance decays, shown in Fig. \ref{fig:timedep} and
Tab.~\ref{timediffs}.  

The correlation functions have been fitted using the numerical
Monte-Carlo procedure described in Sect.~\ref{sec:Therm}. Fig. \ref{fig:pikpfun} shows
the resulting best-fit functions as solid lines. The fit is performed
simultaneously to the positive $k^{*}_{out}$ and negative
$k^{*}_{out}$ part of the correlation function. The lines in
Fig. \ref{fig:pikpdr} are simply the two parts of the fit function from
Fig. \ref{fig:pikpfun} divided by each other, they are not fit
independently. 

\begin{figure}[!htb1]
\begin{center}
\includegraphics*[angle=0, width=7cm]{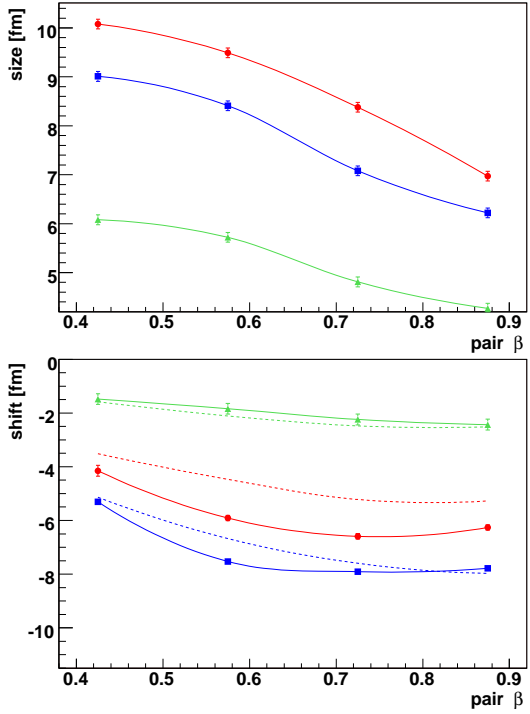}
\end{center}
\caption{\emph{ \small Parameters of the non-identical particle
sources from fits. In the upper panel source size is shown, in the lower - shift
between mean emission points in the direction of pair transverse
momentum. Red circles are for pion-kaon, Blue squares for pion-proton,
green triangles for kaon-proton. Dashed lines are input overall
asymmetries from Fig.~\ref{fig:avcomp}}}
\label{fig:nonidsum}
\end{figure}

The results of the fit are shown in Fig. \ref{fig:nonidsum}. We find
that the size of the emitting system decreases with pair velocity for
all considered pair types. This is consistent with the ``$m_{T}$
scaling'' of the HBT radii observed in the identical particle
femtoscopy calculations. The shifts between various particle species
have an expected ordering - the larger the mass difference, the larger
the shift. This means that for all observed systems the lighter
particle is emitted, on the average, closer to the center and/or later
than the heavier one. It also means that time differences do not
change the qualitative behaviour of the observed asymmetries. However,
consulting Tab.~\ref{timediffs} and Fig.~\ref{fig:avcomp}, one can see
that they do contribute to the observed asymmetry. The dashed lines
in Fig.~\ref{fig:nonidsum} show the overall asymmetry from
Fig.~\ref{fig:avcomp} as predicted
directly from the emission functions $S_{1,2}(\bold r)$ from the model. The agreement in
absolute values and general trends between them and the final fit
values is within $1.0 fm$. 

This is of crucial importance for the experiment. It means that by
performing an advanced fit of all three combinations of non-identical
particle correlation functions one can indeed infer the properties of
the underlying two-particle emission functions, and therefore obtain a
new, unique piece of information about the dynamics of the collision.

\begin{figure}[!htb1]
\begin{center}
\includegraphics*[angle=0, width=7cm]{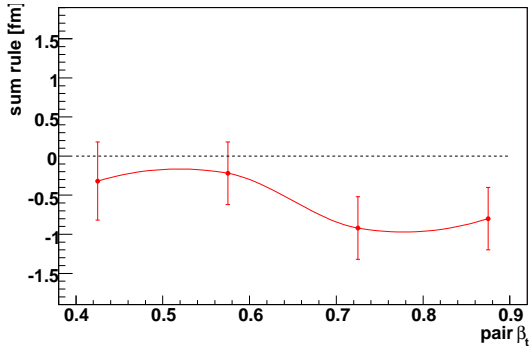}
\end{center}
\caption{\emph{ \small Test of the sum rule of mean shifts between
various particle types.}}
\label{fig:sumrule}
\end{figure}

One can also test the validity of the sum rule (\ref{sumform}). The
test using the fit values from Fig.~\ref{fig:nonidsum} is
shown in Fig.~\ref{fig:sumrule}. One can see that the rule is valid
for smaller pair velocity and holds only approximately for larger
velocities. This should be compared to results in
Tab.~\ref{spacediffs}, where the shifts obtained from the separation
distributions themselves follow the sum rule with the accuracy of $0.1
fm$. Also one can expect deviations on the order of the differences
between the input and fitted values of the shift shown in
Fig.~\ref{fig:nonidsum}. One can see that within these systematic
limits the agreement is acceptable. This is another important
conclusion for the experiment. If shifts between all three
combinations are measured in the comparable pair velocity window one
expects the sum rule for the shifts (\ref{sumform}) to hold. It can be
used as a quality check on the data.

\begin{figure}[!htb1]
\begin{center}
\includegraphics*[angle=0, width=7cm]{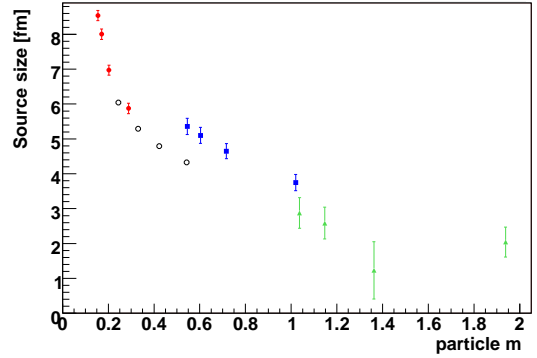}
\end{center}
\caption{\emph{ \small Single particle source sizes obtained from
non-identical particle correlation fits, according
to~(\ref{onefromtwo}). Red circles are pions sizes, blue squares -
kaon, green triangles - proton. Open circles are sizes obtained from
identical-particle $\pi$ correlations.}}
\label{fig:singler}
\end{figure}

One can also use the fitted two-particle radii for all systems to obtain the single
particle radii according to~(\ref{onefromtwo}). The results are shown
in Fig.~\ref{fig:singler}. The radii are reasonable and exhibit the
expected ``$m_{T}$ scaling'' for all particle species. Therefore an
experiment can extract the information not only about the asymmetries
of emission but also about the size of the system. 
The comparison between the size estimates obtained from non-identical
particle correlations and ``HBT radii'' obtained from identical pion
interferometry was done. Both sizes for pions are consistent with each other. 
One must remember that in the case of Therminator model, the obtained
source functions exhibit large long-range non-gaussian tails~\cite{Kisiel:2006is}. On the
other hand, both identical and non-identical femtoscopic sizes were
obtained assuming perfect gaussian source. Both of these measures can
be sensitive to long-range tails in a different way, so the
comparison must be done with caution.

\section{Summary}

We have presented the first complete set of calculations of
non-identical particle correlations from the single freeze-out models
with complete treatment of resonances. The non-identical particle
femtoscopy method was shown to be sensitive to both the size and
emission asymmetries in the system. The observed effects have been
shown to be under control both qualitatively and quantitatively. The
method to extract femtoscopic information from such correlation
functions have been presented and employed to the
``pseudo-experimental'' functions obtained from the model. The results
of the fit have been shown to be in agreement with the characteristics of the
input source. Several consistency checks on the experimental data have
been proposed and their validity tested. A method to test the
consistency of identical and non-identical femtoscopy results has
also been proposed. 

The femtoscopic analysis of non-identical particle correlations has
shown that the emission asymmetries between pions, kaons and protons
are expected to occur in heavy-ion collisions. Spatial asymmetry coming
from radial flow was observed. Time asymmetry coming from resonance
propagation and decay was also estimated and found to be on the order
of the space asymmetry in the laboratory frame and in the same
direction for all tested systems. Therefore a realistic and
self-consistent estimate of the two effects has been given for the
first time. Consistency between identical and non-identical
femtoscopic sizes was tested and found to be within 1.0 fm.

Predictions for both the size of the system as well as emission
asymmetry has been given for the central AuAu collisions for
pion-kaon, pion-proton and kaon-proton correlations. 

\bigskip
\noindent\textbf{Acknowledgements}
\medskip

This work was supported by Polish Ministry of Science and Higher
Education, grants no. 0395/P03/2005/29 and 134/E-365/SPB/CERN/P-03/DWM
97/2004-2007. I would like to thank
prof. W. Broniowski and prof. W. Florkowski for
very fruitful discussions.



\begin{thebibliography}{99}


\bibitem{Florkowski:2001fp}
  W.~Florkowski, W.~Broniowski and M.~Michalec,
  ``Thermal analysis of particle ratios and p(T) spectra at RHIC,''
  Acta Phys.\ Polon.\ B {\bf 33}, 761 (2002)
  [arXiv:nucl-th/0106009].

\bibitem{Broniowski:2001we}
  W.~Broniowski and W.~Florkowski,
  ``Explanation of the RHIC p(T)-spectra in a thermal model with expansion,''
  Phys.\ Rev.\ Lett.\  {\bf 87}, 272302 (2001)
  [arXiv:nucl-th/0106050].

\bibitem{Broniowski:2002wp}
  W.~Broniowski, A.~Baran and W.~Florkowski,
  ``Thermal model at RHIC. II: Elliptic flow and HBT radii,''
  AIP Conf.\ Proc.\  {\bf 660}, 185 (2003)
  [arXiv:nucl-th/0212053].

\bibitem{Kisiel:2006is}
  A.~Kisiel, W.~Florkowski and W.~Broniowski,
  ``Femtoscopy in hydro-inspired models with resonances,''
  Phys.\ Rev.\ C {\bf 73}, 064902 (2006)
  [arXiv:nucl-th/0602039].

\bibitem{Adams:2003qa}
  J.~Adams {\it et al.}  [STAR Collaboration],
  ``Pion kaon correlations in Au + Au collisions at s(NN)**(1/2) = 130-GeV,''
  Phys.\ Rev.\ Lett.\  {\bf 91}, 262302 (2003)
  [arXiv:nucl-ex/0307025].

\bibitem{LisaRetiere} 
  F.~Retiere and M.~A.~Lisa,
  ``Observable implications of geometrical and dynamical aspects of  freeze-out
  in heavy ion collisions,''
  Phys.\ Rev.\ C {\bf 70}, 044907 (2004)
  [arXiv:nucl-th/0312024].

\bibitem{Lednicky:1982}
  R.~Lednicky, V.L.~Lyuboshitz,
  Yad.\ Fiz.\ {\bf 35} (1982) 1316 (Sov.\ J.\ Nucl.\ Phys.\ {\bf 35}
(1982) 770); Proc.\ Int.\ Workshop on Paricle Correlations and
Interferometry in Nuclear Collisions, CORINNE 90, Nantes, France 1990
(ed. D.~Ardouin, World Scientific 1990) p. 42; Heavy Ion Physics {\bf
3} (1996) 1

\bibitem{Lednicky:1995vk}
  R.~Lednicky, V.~L.~Lyuboshits, B.~Erazmus and D.~Nouais,
  ``How to measure which sort of particles was emitted earlier and which
  later,''
  Phys.\ Lett.\ B {\bf 373}, 30 (1996).

\bibitem{Voloshin:1997jh}
  S.~Voloshin, R.~Lednicky, S.~Panitkin and N.~Xu,
  ``Relative space-time asymmetries in pion and nucleon production in
  noncentral nucleus nucleus collisions at high energies,''
  Phys.\ Rev.\ Lett.\  {\bf 79}, 4766 (1997)
  [arXiv:nucl-th/9708044].

\bibitem{Lednicky:2005tb}
  R.~Lednicky,
  ``Finite-size effects on two-particle production in continuous and  discrete
  spectrum,''
  arXiv:nucl-th/0501065.

\bibitem{Kisiel:2006si}
  A.~Kisiel,
  ``Non-identical particle femtoscopy in heavy-ion collisions,''
  AIP Conf.\ Proc.\  {\bf 828}, 603 (2006).

\bibitem{Kisiel:2004phd}
  A.~Kisiel,
  ``Studies of non-identical meson-meson correlation at low relative
velocities in relativistic heavy-ion collisions registered in the STAR
experiment'',
  PhD Thesis, Warsaw Univeristy of Technology, Dec 2004

\bibitem{Kisiel:2005hn}
  A.~Kisiel, T.~Taluc, W.~Broniowski and W.~Florkowski,
  ``THERMINATOR: Thermal heavy-ion generator,''
  Comput.\ Phys.\ Commun.\  {\bf 174}, 669 (2006)
  [arXiv:nucl-th/0504047].

\bibitem{Eidelman:2004wy}
  S.~Eidelman {\it et al.}  [Particle Data Group],
  ``Review of particle physics,''
  Phys.\ Lett.\ B {\bf 592}, 1 (2004).

\bibitem{Adams:2003qm}
  J.~Adams {\it et al.}  [STAR Collaboration],
  ``Pion, kaon, proton and anti-proton transverse momentum distributions  from
  p + p and d + Au collisions at s(NN)**1/2 = 200-GeV,''
  Phys.\ Lett.\ B {\bf 616}, 8 (2005)
  [arXiv:nucl-ex/0309012].

\bibitem{Kisiel:Nukl}
A.~Kisiel,
``CorrFit - a program to fit arbitrary two-particle correlation
functions'' 
NUKLEONIKA 2004, {\bf 49(Supplement 2)}:s81-s83

\end{thebibliography}
\end{document}